\documentclass[%
reprint,
amsmath,amssymb,
aps,
prb,
]{revtex4-2}

\usepackage{graphicx}
\usepackage{dcolumn}
\usepackage{bm}
\usepackage{appendix}

\usepackage{color}
\usepackage{braket}

\begin{document}

\preprint{APS/123-QED}

\title{
Quantum--classical correspondence and dissipative to dissipationless crossover in magnetotransport phenomena}

\author{Akiyoshi Yamada$^1$}
\author{Yuki Fuseya$^{2, 3}$}%
\affiliation{%
$^1$The Institute for Solid State Physics, the University of Tokyo, Chiba 277-8581, Japan\\
$^2$Department of Engineering Science, University of Electro-Communications, Chofu, Tokyo 182-8585, Japan\\
$^3$Institute for Advanced Science, University of Electro-Communications, Chofu, Tokyo 182-8585, Japan
}%




\date{\today}

\begin{abstract}
The three-dimensional magneto-conductivity tensor was derived in a gauge invariant form based on the Kubo formula considering the quantum effect under a magnetic field, such as the Landau quantization and the quantum oscillations.
We analytically demonstrated that the quantum formula of the magneto-conductivity can be obtained by adding a quantum oscillation factor to the classical formula.
This result establishes the quantum--classical correspondence, which has long been missing in magnetotransport phenomena.
Moreover, we found dissipative-to-dissipationless crossover in the Hall conductivity by paying special attention to the analytic properties of thermal Green's function.
Finally, by calculating the magnetoresistance of semimetals, we identified a phase shift in quantum oscillation originating from the dissipationless transport predominant at high fields. 
\end{abstract}

\maketitle

\section{Introduction}
Magnetotransport phenomena constitute one of the oldest research topics in solid state physics~\cite{Thomson1857,Kapitza1928,Pippard_book,Beer1963}.
In particular, applying a magnetic field can drastically alter the electron transport through the Lorentz force.
From the classical equation of motion, the magneto-conductivity tensor can be expressed in the following form~\cite{grosso2000solid,Zhu2018}:
\begin{eqnarray}
\hat{\sigma}_{\rm cl}&=&
\sigma_0 \left( \begin {array}{ccc}
\displaystyle\frac{1}{(\omega_c\tau)^2+1} & \displaystyle\frac{-\omega_c\tau }{(\omega_c\tau)^2+1} & 0\\[15pt]
\displaystyle \frac{\omega_c\tau }{(\omega_c\tau)^2+1} & \displaystyle \frac{1}{(\omega_c\tau)^2+1} & 0\\ 
\noalign{\medskip}0 & 0 & 1 \end {array} \right),
\label{Eq.1}
\end{eqnarray}
where $\sigma_0$ is the conductivity at zero magnetic field, $\omega_c$ is the cyclotron frequency, and $\tau$ is the relaxation time.
A similar equation can be derived based on the semiclassical Boltzmann's equation~\cite{Wilson2011-bx}.
The physical meaning of the classical Eq. \eqref{Eq.1} is clear and straightforward to handle. Thus, Eq. \eqref{Eq.1} has been a powerful tool for analyzing the transport properties in good metals with a larger chemical potential than the cyclotron energy, $ \mu \gg \hbar \omega_c$.
On the other hand, the validity of the (semi-) classical theory is lost when $\hbar \omega_c \gtrsim \mu$ or $\omega_c \tau \gtrsim 1$, where the quantum effect, such as the Landau quantization, plays a crucial role.

The magnetotransport phenomena have recently attracted renewed interest, especially in topological materials~\cite{Novoselov2005,Fuseya2015JPSJ,Feng2015,Shekhar2015,Luo2015,Wang2018}.
The effect of Landau quantization cannot be neglected in these materials, even for magnetic fields of several teslas, because most of them have a small effective mass yielding high cyclotron energy.
Nevertheless, researchers relied only on the (semi-) classical formula to analyze the experimental data of topological materials because the complexity of applying the previous quantum formula to actual materials is high except for a few simple cases \cite{lu2015prb,wang2016prl,Konye2018prb}.
The lack of understanding of the connection between the quantum and the classical formula further prevents us from analyzing the experimental data from the quantum perspective.

The quantum counterpart of the classical formula can be obtained based on the Kubo formula~\cite{Kubo1957,KUBO1965}.
For two-dimensional systems, the Kubo formula has successfully elucidated the underlying physics of the quantum Hall effect~\cite{Ando1982_review,Dmitriev2003,Dmitriev_review_2012}.
Moreover, since the late 1950s, the fundamentals of magnetotransport phenomena has been investigated for three-dimensional (3D) system~\cite{Argyres1958,ADAMS1959,KUBO1965,Fukuyama1970MR,Shiba1971}, among which certain theoretical investigations followed zero-temperature formulation~\cite{Argyres1958,ADAMS1959}.

Kubo {\it et al.} showed that the quantum formula agrees with the classical one, Eq. \eqref{Eq.1}, in the weak magnetic field limit, but the quantum--classical correspondence could not be obtained in the strong field, where the effects of the Landau quantization and the quantum oscillation are prominent~\cite{KUBO1965}.
Abrikosov derived another quantum formula valid only in the strong field region so that the quantum--classical correspondence could not be obtained.
More elegant formulations were established using the Kubo formula and the thermal Green's function by Fukuyama {\it et al.}~\cite{Fukuyama1970MR} and Shiba {\it et al.}~\cite{Shiba1971}.
Although Shiba's formulations are valid in the entire magnetic field region (from weak to strong), the obtained formula was too complicated to uncover the clear quantum--classical correspondence.

In this study, we revisit the quantum magnetotransport of 3D free electrons based on the Kubo formula~\cite{Kubo1957,AGD}.
We succeeded in obtaining a simple quantum formula, Eq. \eqref{Q-C_correspondence}, by paying special attention to the analytic nature of Green's function and putting some originality into the derivation.
The developed formula is essentially equivalent to Abvikosov's formula at strong fields and to the theory by 
Shiba {\it et al.} for the entire region, but in a much simpler form.
The quantum--classical correspondence is clear in the developed formula: just add the quantum oscillation factor $Q(B)$ in the diagonal conductivity $\sigma_{xx,yy}$.
Furthermore, as a by-product, we found that the crossover from the dissipative term to the dissipationless one occurs in the Hall conductivity $\sigma_{xy}$ by increasing the magnetic field, whereas the diagonal conductivity $\sigma_{xx}$ are dominated by the dissipative term for the entire range of the magnetic field.

\begin{figure}[t]
\begin{center}
	\includegraphics[width=8.5cm]{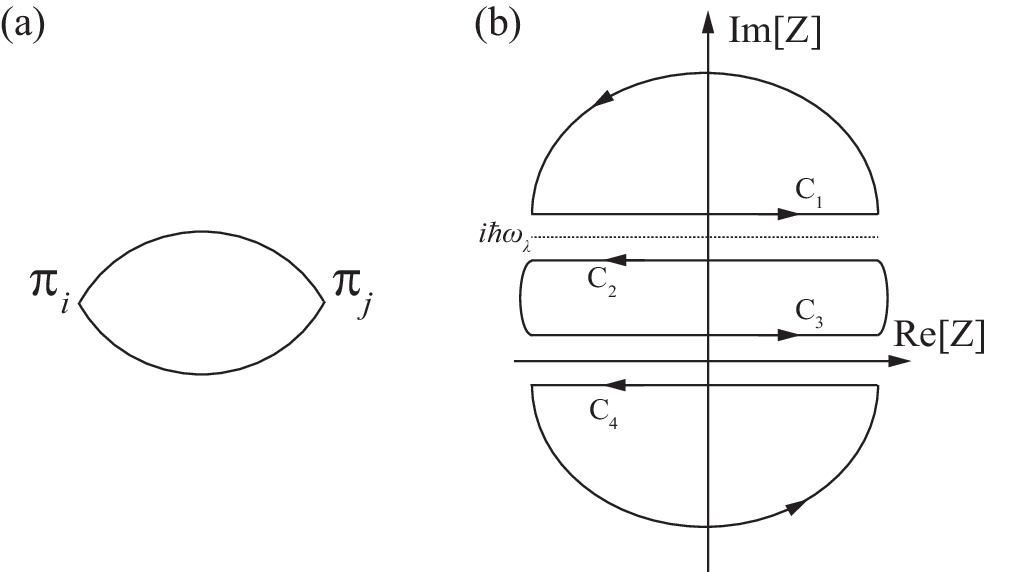}
	\caption{\label{fig1} (a) Feynman diagram for conductivity tensor in a magnetic field. (b) Path of integration in the summation of Matsubara frequency.}
\end{center}
\end{figure}

\section{Theory}
\subsection{Magnetotransport theory based on the Kubo formula and Green's function}
The components of the conductivity tensor were evaluated considering the Kubo formula in a magnetic field~\cite{Kubo1957,KUBO1965,Fukuyama1970MR,Shiba1971,Mahan2000,Fuseya2015JPSJ}, as follows:
\begin{eqnarray}
\sigma_{ij}&=&\frac{1}{i}\left.\frac{\partial \Phi_{ij}}{\partial\omega}\right|_{\omega=0},\label{Kubo1} \\
\Phi_{ij} &=& -\frac{2e^2}{\beta V m^2} \sum_{n,{\bm k}} {\rm Tr}\left[ \mathcal{G}\pi_i \mathcal{G}\pi_j\right],
\end{eqnarray}
where $\Phi$ denotes the current-current correlation function [Fig. \ref{fig1} (a)], $\pi_i$ indicates the kinematical momentum operator, and $\mathcal{G}=(i\varepsilon_n-\mathcal{H})^{-1}$ represents the thermal Green's function ($\varepsilon_n$ denotes the Matsubara frequency) \cite{Matsubara1955}. 
$\mathcal{H}$ denotes the Hamiltonian of a 3D free electron in the magnetic field.
$\beta=1/k_BT$, and $V(=L^3)$ denotes the system volume.
In the magnetic field, the velocity operator is given by $\pi_i/m=(p_i+eA_i)/m$, where the electron charge is defined as $-e<0$.
Considering the trace over the Landau indices $\ell,\ell'$, the correlation function can be rewritten in the following form:
\begin{eqnarray}
\Phi_{ij}(i\omega_\lambda) &=& -\frac{2e^2N_L}{Lm^2}\sum_{\ell,\ell'}\braket{\ell|\pi_i|\ell'}\braket{\ell'|\pi_j|\ell}F_{\ell,\ell'}(i\omega_\lambda),\nonumber\\
F_{\ell,\ell'}(i\omega_\lambda)&=&\frac{1}{\beta}\sum_{n,{ k_z}}g_{\ell'}(k_z,i\varepsilon_n)g_{\ell}(k_z,i\varepsilon_n-i\omega_\lambda),\nonumber
\end{eqnarray}
where $g_\ell=(i\varepsilon_n-E_\ell)^{-1}$ and $E_\ell=(\ell+\frac{1}{2})\hbar\omega_c+\hbar^2k_z^2 /2m$.
$N_L(=eB/2\pi\hbar)$ is the Landau degeneracy \cite{grosso2000solid,Fuseya2015JPSJ}.
$\omega_c=eB/m$ denotes the cyclotron frequency, where $B$ denotes the magnetic field.
With the magnetic field along the $z$-axis, the $\pi$ operators in the $x$-$y$ plane can be defined as $\pi_x=\sqrt{\hbar eB/2}(a^++a^-)$, $\pi_y=i\sqrt{\hbar eB/2}(-a^++a^-)$, where $a^{+}$ and $a^{-}$ represent the raising and lowering operators for Landau indices, respectively.
From the commutation relation of the kinematical momentum in a magnetic field, $[\pi_i, \pi_j]=-i\hbar e \varepsilon_{ijk}B_k$, a commutation between $\pi_x \leftrightarrow \pi_y$ implies the sign inversion of the magnetic field, which guarantees the Onsager's reciprocal relation $\sigma_{xy}(-B)=\sigma_{yx}(B)$.
Furthermore, we do not need to assume the gauge by transfer from the vector potential to the magnetic field through the commutation relation. Thus, the formulas obtained below are all gauge invariant.

\begin{figure*}[t]
\begin{center}
	\includegraphics[width=16cm]{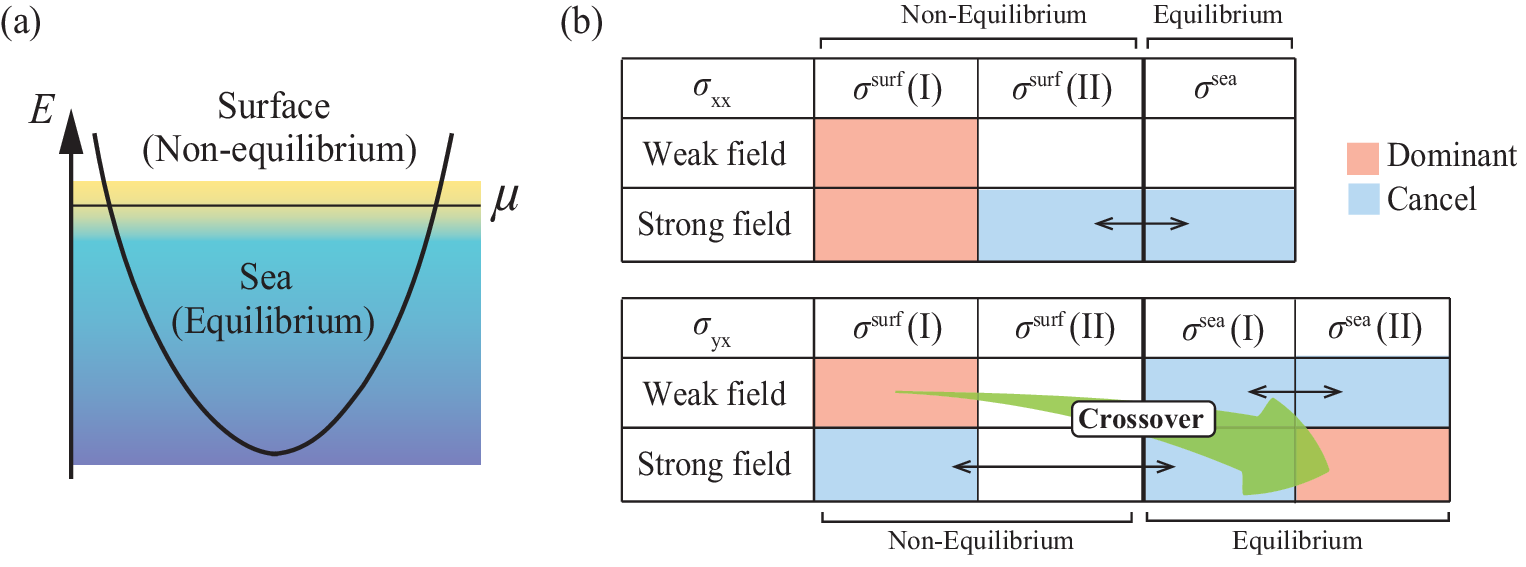}
	\caption{\label{fig2} (a) Energy dispersion and schematic of the electron distribution in the Fermi surface and Fermi sea. (b) Table listing the dominant terms of the magnetoconductivity tensor in the weak and strong fields.}
\end{center}
\end{figure*}

We can conduct the summation over $\ell'$ and obtain the following form:
\begin{eqnarray}
\Phi_{xx}(i\omega_\lambda) &=& -\frac{e^3\hbar B}{Lm^2}\sum_{\ell}\left[\ell F_{\ell,\ell-1}+(\ell+1)F_{\ell,\ell+1}\right], \label{eq_corr_func_xx}\\
\Phi_{yx}(i\omega_\lambda) &=& i\frac{e^3\hbar B}{Lm^2}\sum_{\ell}\left[\ell F_{\ell,\ell-1}-(\ell+1)F_{\ell,\ell+1}\right].\label{eq_corr_func_xy} 
\end{eqnarray}
The summation with respect to $\varepsilon_n$ can be rewritten into the path integration along the imaginary axis in the complex plane by introducing the Fermi distribution function $n_F(x)=1/(e^{\beta(x-\mu)}+1)$~\cite{AGD,Fuseya2015JPSJ}.
Furthermore, the pass of the integral was transformed into four separate improper integrals along the real axis to avoid crossing the singularities of Green's functions [${\rm Im}[z]=0$, ${\rm Im}[z]=\omega_\lambda$; cf. Fig. \ref{fig1} (b)].
After an analytic continuation, $i\omega_\lambda \rightarrow \hbar\omega+i \delta$,
$F_{\ell',\ell}$ can be expressed as follows:
\begin{widetext}
\begin{eqnarray}
	F_{\ell,\ell'}=-\frac{1}{2\pi i}\sum_{k_z}\int_{-\infty}^{\infty}dx \,n_F(x)\left[G_{\ell'}^R(x+\hbar\omega)G_{\ell}^R(x) -G_{\ell'}^R(x+\hbar\omega)G_{\ell}^A(x) \right.\nonumber \\
	\left. +G_{\ell'}^R(x)G_{\ell}^A(x-\hbar\omega) -G_{\ell'}^A(x)G_{\ell}^A(x-\hbar\omega) \right],\nonumber
\end{eqnarray}
\end{widetext}
where $G^{A(R)}_\ell(x)=(x+E_\ell\mp i\Gamma)^{-1}$ represents the advanced (retarded) Green's function, respectively.
We introduced the imaginary part of the self-energy as $\Gamma=\hbar/2\tau$.
The second and third terms, including $G^A G^R$ and $G^R G^A$, are referred to as the ``Fermi surface terms," accounting for the contribution from the non-equilibrium transport in the vicinity of the Fermi energy, whereas the first and fourth terms with $G^RG^R$ ($G^AG^A$) are referred to as the `` Fermi sea terms" corresponding to the equilibrium transport [cf. Fig. \ref{fig2} (a)].

\subsection{Fermi surface terms}
We first investigated the contributions from the Fermi surface terms at low temperatures.
The Fermi surface terms can be expressed proportionally to $-\partial n_F / \partial x$, which can be approximated by a delta function at low temperatures.

The transverse components of the magnetoconductivity tensor, $\sigma_{xx, yy}$ and $\sigma_{xy, yx}$, were calculated using Eqs.(\ref{Kubo1}), (\ref{eq_corr_func_xx}), and (\ref{eq_corr_func_xy}).
After integrating with respect to $k_z$, the following equation can be obtained:
\begin{eqnarray}
    \sigma_{xx}&=&\sigma_{xx}^{\rm surf}({\rm I})+\sigma_{xx}^{\rm surf}({\rm II}),\nonumber\\
\sigma_{yx}&=&\sigma_{yx}^{\rm surf}({\rm I})+\sigma_{yx}^{\rm surf}({\rm II}),\nonumber
\end{eqnarray}
where
\begin{widetext}
\begin{eqnarray}
\sigma_{xx}^{\rm surf}({\rm I})&=&\sigma_0\frac{1}{(\omega_c\tau)^2+1}\sum_{\ell} 3(\gamma\omega_c\tau)^2{\rm Re}\left[\frac{\ell+\frac{1}{2}}{K_\ell}\right],\quad 
\sigma_{xx}^{\rm surf}({\rm II})=-\sigma_0\frac{1}{(\omega_c\tau)^2+1}\sum_{\ell}\frac{3\gamma^2(\omega_c\tau)^3}{2} {\rm Im}\left[\frac{1}{K_\ell}\right], \label{eq_sigma_xx_surf}\\
\sigma_{yx}^{\rm surf}({\rm I})&=&\sigma_0\frac{ \omega_c\tau}{(\omega_c\tau)^2+1}\sum_{\ell}3(\gamma\omega_c\tau)^2{\rm Re}\left[\frac{\ell+\frac{1}{2}}{K_\ell}\right],\quad 
\sigma_{yx}^{\rm surf}({\rm II})=\sigma_0\frac{1}{(\omega_c\tau)^2+1}\sum_{\ell} \frac{3(\gamma\omega_c\tau)^2}{2}{\rm Im}\left[\frac{1}{K_\ell}\right], \label{eq_sigma_yx_surf}\\
K_\ell&=&\sqrt{1-(2\ell+1)\omega_c\tau\gamma+i\gamma}\nonumber
\end{eqnarray}
\end{widetext}
$\sigma_0$ is the electric conductivity in zero-field given by $\frac{e^2\tau}{m}\frac{1}{3\pi^2}\left({\sqrt{2m\mu}}/{\hbar}\right)^3$ and $\gamma={\Gamma}/{\mu}$.
During the derivation of the above form, we transformed the equation to be compatible with the classical formula, say, by factorizing $1/[(\omega_c \tau)^2 + 1]$.

\begin{figure}[tb]
\begin{center}
	\includegraphics[width=7cm]{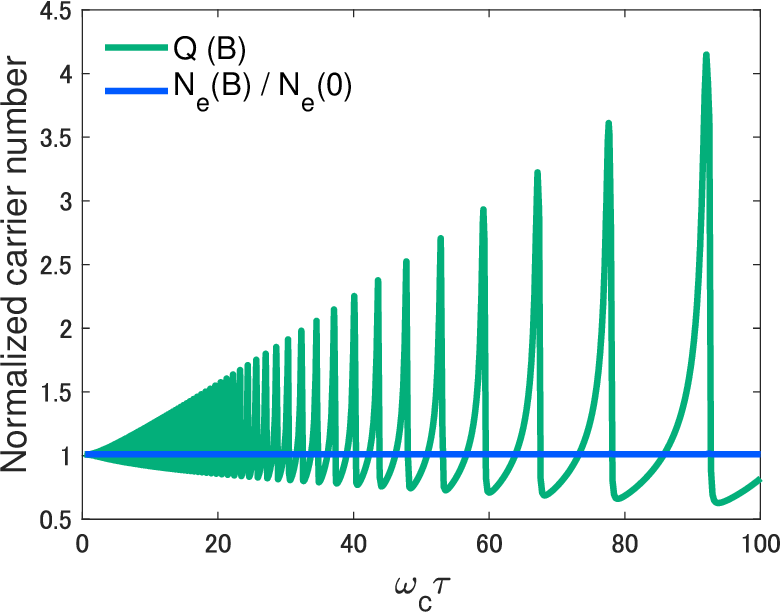}
	\caption{\label{figC} Field dependence of $Q(B)$ and $N_e (B)$ for $\mu =10$ meV at zero-field and $\Gamma=0.01$ meV. $Q(B)$ exhibits a clear quantum oscillation, providing a quantum correction to the classical magnetotransport.}
\end{center}
\end{figure}

\begin{figure*}[tb]
\begin{center}
	\includegraphics[width=15cm]{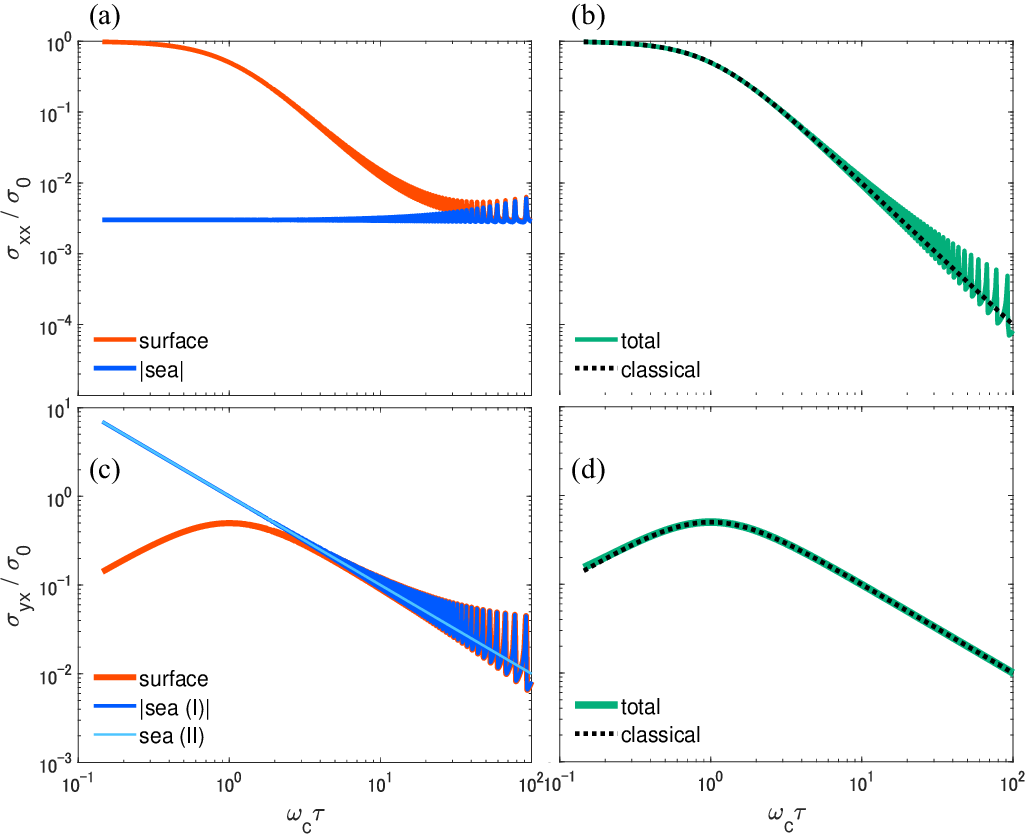}
	\caption{\label{fig3} Field dependence of (a) transverse and (c) Hall conductivity from the surface and sea terms.
		(b,d) Field dependence of total conductivity and classical magnetoconductivity.
		$\mu =10$ meV at zero-field and $\Gamma=0.01$ meV.}
\end{center}
\end{figure*}

The longitudinal component of the correlation function can be expressed as follows:
\begin{eqnarray}
\Phi_{zz}(i\omega_\lambda) =\frac{-2e^2}{\beta Lm^2}\sum_{\ell,n,{k_z}}p_z^2g_{\ell}(k_z,i\varepsilon_n)g_{\ell}(k_z,i\varepsilon_n-i\omega_\lambda).
\end{eqnarray}
Note that the velocity operator along the magnetic field ($z$ axis) contains no transition component across different Landau indices.
Then, the longitudinal conductivity from the Fermi surface term is obtained as follows:
\begin{eqnarray}
\sigma_{zz} &=& \frac{e^2\tau}{m} N_e(B),\label{eq_sigma_zz}\\
N_e(B)&=&\left(\frac{\sqrt{2m\mu}}{\hbar}\right)^3\frac{\gamma\omega_c\tau}{\pi^2}\sum_{\ell}{\rm Re}[{K_\ell}]. \nonumber
\end{eqnarray}
where, $N_e$ is independent from the magnetic field and equivalent to the carrier number for $\Gamma \ll \mu$, as shown in Fig. \ref{figC}.

The remaining components, such as $\sigma_{xz}$ and $\sigma_{yz}$, are zero due to the relation:
\begin{eqnarray}
\braket{\ell|\pi_{x,y}|\ell'}\braket{\ell'|\pi_z|\ell} &=&(C_\ell\delta_{\ell'-1,\ell}+C'_\ell\delta_{\ell'+1,\ell})\delta_{\ell',\ell}
\nonumber\\
&=&0.
\end{eqnarray}
(Note that $\pi_x$ and $\pi_y$ are off-diagonal, and $\pi_z$ is the diagonal based on the Landau indices.)

\subsection{Fermi sea terms}
The Fermi sea terms, consisting of $G^RG^R$ or $G^AG^A$, include the contribution from deep energy states due to a factor of $n_F$, which is the origin of the dissipationless contribution.
These terms can be transformed into integrating the derivative of $n_F$ after performing integration by part to obtain their analytical forms, which turned out to be a key derivative to obtain the clear quantum--classical correspondence in the end.

The diagonal component of the transverse conductivity tensor contains only a single term from the sea term as follows:
\begin{eqnarray}
\sigma_{xx}^{\rm sea} &=&\sigma_0\sum_\ell \frac{3\gamma^2\omega_c\tau}{2}{\rm Im}\left[\frac{1}{K_\ell}\right]
\end{eqnarray}
Unexpectedly, this form perfectly cancels one of the Fermi surface terms, $\sigma_{xx}^{\rm surf}({\rm II})$ in Eq.(\ref{eq_sigma_xx_surf}), at strong fields as \[ \sigma_{xx}^{\rm sea}+\sigma_{xx}^{\rm surf}({\rm II})=0
\quad ({\rm for}\,\, \omega_c \tau \gg 1).\]
Therefore, only the dissipative term $\sigma_{xx}^{\rm surf} ({\rm I})$ contributes to the diagonal conductivity, and no dissipationless term remains.

The Hall conductivity includes two sea terms:
\begin{eqnarray}
\sigma_{yx}^{\rm sea}({\rm I}) &=& \sigma_0\sum_\ell 3\gamma^2\omega_c\tau{\rm Re}\left[\frac{\ell+\frac{1}{2}}{K_\ell}\right]\label{eq_sigma_xy_sea}\\
\sigma_{yx}^{\rm sea}({\rm II}) &=& \sigma_0\sum_\ell 3\gamma{\rm Re}\left[K_\ell\right].
\end{eqnarray}
Contrastingly, at low fields, these terms cancel each other as
\[ \sigma_{yx}^{\rm sea} ({\rm I}) + \sigma_{yx}^{\rm sea}({\rm II}) =0
\quad ({\rm for}\,\, \omega_c \tau \ll 1 \text{ and } \Gamma\ll\mu),\]
At high fields, on the other hand, the first term $\sigma_{yx}^{\rm sea}({\rm I})$ neutralizes $\sigma_{yx}^{\rm surf}({\rm I})$ in Eq.(\ref{eq_sigma_yx_surf}) as
\[ \sigma_{yx}^{\rm surf}({\rm I}) + \sigma_{yx}^{\rm sea}({\rm I}) =0 
\quad ({\rm for}\,\, \omega_c \tau \gg 1).\]
All cancellation relations are summarized in Fig. \ref{fig2} (b).

The second term $\sigma_{yx}^{\rm sea}({\rm II})$ can be expressed as follows:
\begin{eqnarray}
\sigma_{yx}^{\rm sea}({\rm II}) &=& \frac{eN_e(B)}{B},
\end{eqnarray}
which is independent of scattering as it hardly depends on $\tau$.
This aspect is consistent with the classical description of the Hall effect, which is reproduced by Kubo~\cite{Kubo1957} for weak field and introduced phenomenologically by Shiba {\it et al.} \cite{Shiba1971} and Ando {\it et al.} \cite{Ando_Hall_1975}.
The sea term in $\sigma_{zz}$ was evaluated as follows:
\begin{eqnarray}
\sigma_{zz}^{\rm sea} &=&\sigma_0 \sum_\ell 3\gamma^2\omega_c\tau{\rm Im} \left[\frac{1}{K_{\ell}}\right].
\end{eqnarray}
This term is substantially smaller than the surface term if the scattering is weak, $\mu \gg \Gamma$.

\subsection{Quantum--classical correspondence}
Finally, the total conductivity tensor was obtained for $\Gamma \ll \mu$ as follows:
\begin{eqnarray}
\hat{\sigma}&=&
\sigma_0 \left( \begin {array}{ccc}
\displaystyle\frac{Q(B)}{(\omega_c\tau)^2+1} & \displaystyle\frac{-\omega_c\tau }{(\omega_c\tau)^2+1} & 0\\[15pt]
\displaystyle \frac{\omega_c\tau }{(\omega_c\tau)^2+1} & \displaystyle \frac{Q(B)}{(\omega_c\tau)^2+1} & 0\\ 
\noalign{\medskip}0 & 0 & 1 \end {array} \right), 
\label{Q-C_correspondence}\\
Q(B) &=& 3\left(\gamma\omega_c\tau\right)^2\sum_{\ell} \left(\ell+\frac{1}{2}\right){\rm Re}\left[\frac{1}{K_\ell}\right]. \nonumber
\end{eqnarray}
These results are the counterpart of the classical formulation expressed in Eq. \eqref{Eq.1}.
The field dependence of $Q(B)$ is illustrated in Fig. \ref{figC}.
In particular, $Q(B)$ exhibits a clear quantum oscillation that cannot be considered in the classical formulation.
In other words, the quantum oscillation $Q(B)$ is the only quantum correction to the diagonal conductivity $\sigma_{xx}(B)$.
The global properties of quantum magnetotransport agree with that of classical quantitatively.
Now, Eq.\eqref{Q-C_correspondence} shows the clear quantum--classical correspondence that holds for the entire range of the field, from a weak to a strong magnetic field. 

The field dependence of the Fermi surface term, sea term, and the total conductivity $\sigma^{\rm surface}+\sigma^{\rm sea}$ are illustrated in Fig.\ref{fig3} (a--d), where the chemical potential $\mu$ was varied to maintain the carrier number fixed for the whole range of the magnetic field.
At low field limits, the total conductivity was quantitatively consistent with the classical model (Fig.\ref{fig3} (b),(d)), whereas at high field limits, the Fermi sea term in the transverse conductivity $\sigma_{xx}$ canceled one of the surface terms.
The Fermi surface term dominated $\sigma_{xx}$ across all the field ranges and displayed prominent quantum oscillation at high field limits.
This oscillation is apparently induced by the one-dimensional density of the states in Eq.(\ref{eq_sigma_xx_surf}).
In contrast, the major part of the Fermi surface term in Hall conductivity was canceled by $\sigma_{yx}^{\rm sea}(\rm I)$ (Eq.(\ref{eq_sigma_xy_sea})), and the remaining sea term ($\sigma_{yx}^{\rm sea}({\rm II})$) did not include oscillations if the carrier number was fixed. Thus, the total $\sigma_{yx}$ displayed no oscillation (Fig.\ref{fig3} (d)).
As such, the notable characteristic of Hall conductance is the crossover from the surface term to the sea term.

\subsection{Crossover from dissipative to dissipationless}
\begin{figure}[tb]
\begin{center}
	\includegraphics[width=8cm]{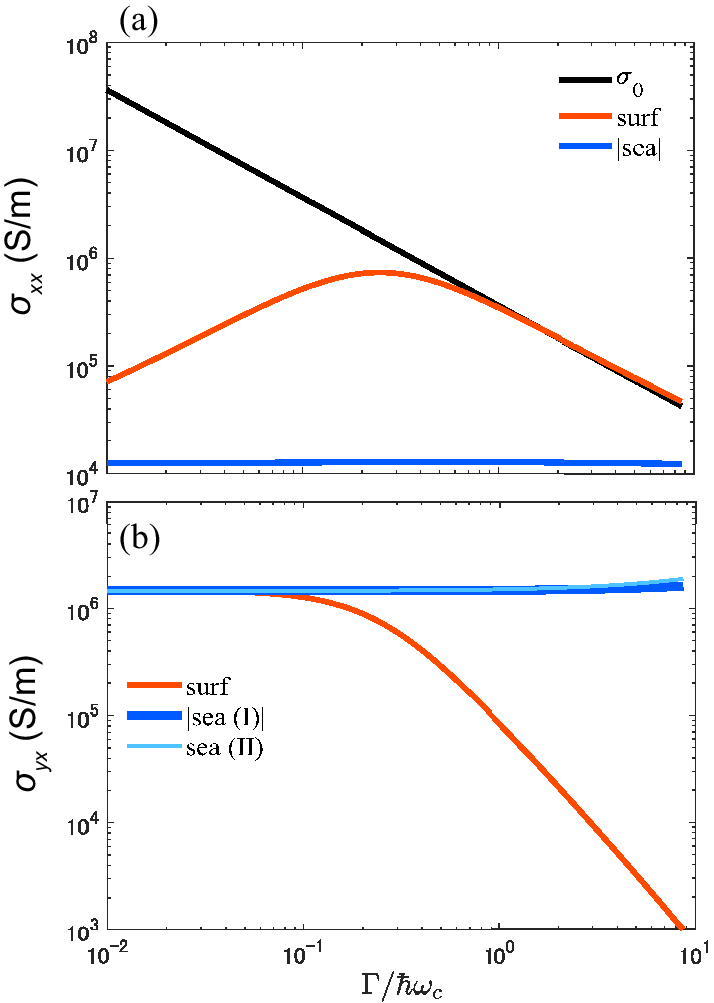}
	\caption{\label{fig4} $\Gamma$ dependence of (a) $\sigma_{xx}$ and (b) $\sigma_{yx}$ at $B=0.5$ T. $\mu=10$ meV at zero-field}
\end{center}
\end{figure}

The physical implication of the ``surface term" and ``sea term" can be clearly understood by examining the $\Gamma$ dependence illustrated in Fig.\ref{fig4}.
The sea terms slightly depend on $\Gamma$, whereas the surface terms strongly depend on it.
This result is expected because the sea terms correspond to the thermodynamic contributions.
The surface terms correspond to the non-equilibrium dissipative transport, while the sea terms correspond to the equilibrium dissipationless transport.
Therefore, the field effect in the Hall conductivity, the crossover from the surface to sea terms, can be regarded as a crossover from dissipative to dissipationless transport.
Note that the surface term in the Hall conductivity gains independence from $\Gamma$ at the clean limit and cancels out with one of the sea terms (Fig.\ref{fig4} (b)).

\section{Shubnikov-de Haas Oscillation in semimetals}
\begin{figure}[t]
\begin{center}
	\includegraphics[width=8cm]{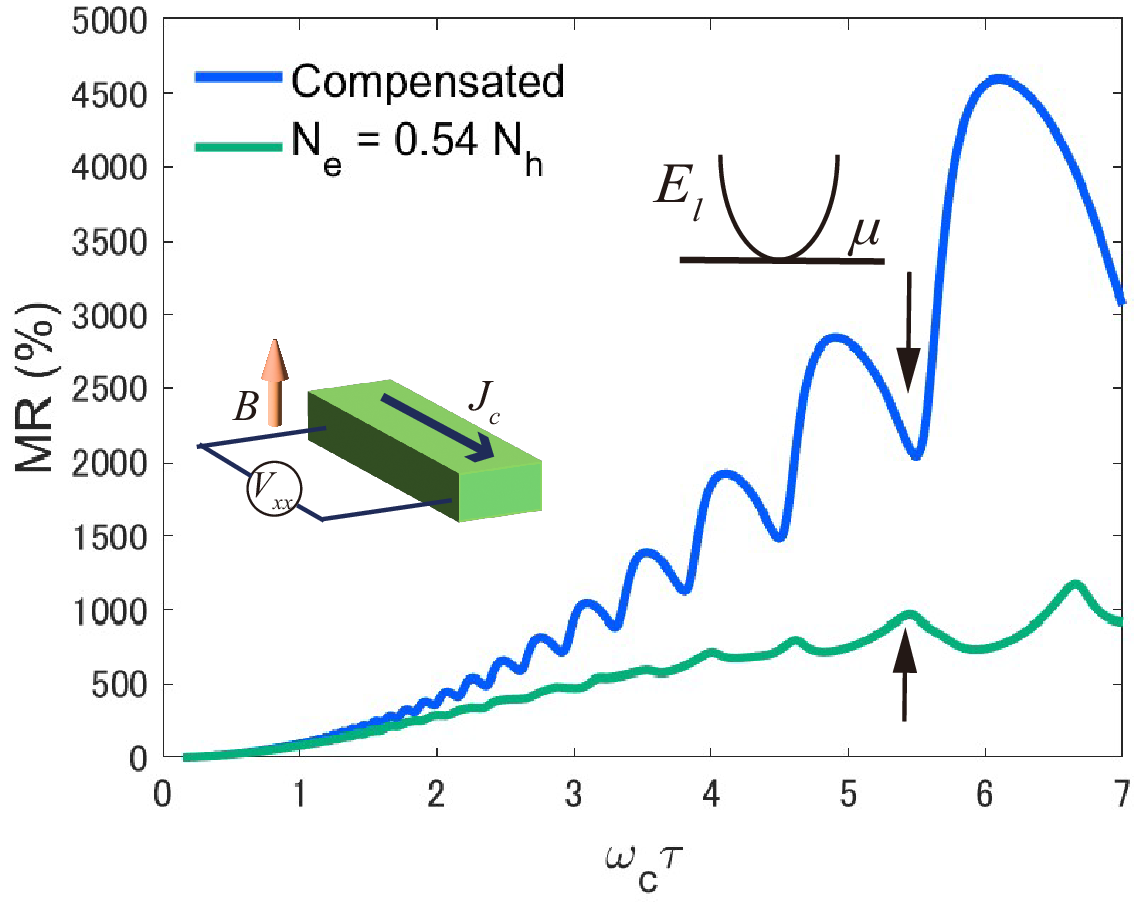}
	\caption{\label{fig5} MR ratio of $\rho_{xx}$ in two-carrier semimetals. $\mu=15$ meV at zero-field and $\Gamma=0.8$ meV.}
\end{center}
\end{figure}

The derived formulation can be readily applied to multicarrier metals and semimetals by adding up the conductivities in each carrier pocket as follows:
\begin{equation}
\hat{\sigma}^{\rm tot} =\sum_i \hat{\sigma}^{\rm (i)}. 
\end{equation}
The magnetoresistivity and Hall resistivity in the voltage measurements were calculated by the inversion of the conductivity tensor:
\begin{equation}
\hat{\rho}=\hat{\sigma}^{-1}.
\end{equation}
As such, a noteworthy remark was found in the calculations for a semimetal.
The transverse magnetoresistance (TMR) in a semimetal with electron and hole carriers is depicted in Fig. \ref{fig5}, where the effective masses of electrons and holes are assumed to be the same.
We calculated the two cases depending on whether the carriers were compensated.
The compensated case exhibited a non-saturating quadratic field dependence across the entire range, whereas the uncompensated case saturated at high field limits.
The quantum oscillations reflected notable variations.
When the Landau level surpassed the chemical potential, the compensated semimetal displayed minimum resistivity, unlike other cases exhibiting the maximum.
The TMR in the isotropic systems can be expressed as follows:
\begin{equation}
\rho_{xx} =\frac{\sigma_{yy}}{\sigma_{xx}\sigma_{yy}-\sigma_{xy}\sigma_{yx}}.
\end{equation}
The field dependence and oscillatory characteristics were determined using the dominant term in the denominator.
If the carrier is compensated, $\sigma_{xy,yx}$ cancels out because these components alter the sign depending on the carrier charge sign.
In this case, the asymptotic form of TMR, is $1/\sigma_{xx}$ at high field limits; therefore, the resistivity does not saturate, and the oscillation is inverted.
In all other cases, $\sigma_{xy,yx}$ dominates the denominator unless they vanish.
Herein, $\rho_{xx}$ approaches $\sigma_{yy}/|\sigma_{xy}|^2$ at a high field limit.
The oscillation is primarily caused by $\sigma_{yy}$ because the oscillation in Hall conductivity is suppressed owing to the cancellation of the dissipative transport.
Furthermore, determining the crosspoints of Landau levels and chemical potential is a critical problem in experiments with semimetals, considering the Berry curvature or the effective $g$-factor in metals and semimetals~\cite{Ando2013,Fuseya2015PRL,Izaki2019PRL}.
The nonnegligible Hall conductivity shifts the position of peaks in $\rho_{xx}$ and may cause misinterpretation of the phase information in the oscillation.
Thus, the present calculation provides a microscopic explanation for this technical problem.

\section{Discussion and Conclusions}
In this study, we derived a formulation regarding magneto-conductivity for 3D free electrons based on the Kubo formula with a Landau-quantized basis, which is valid for an arbitrary field intensity.
We obtained the clear quantum--classical correspondence, as illustrated in Eq. \eqref{Q-C_correspondence}, a long-missing piece in magnetotransport theory.
We found that only the quantum correction to the classical formulation is the quantum oscillation factor $Q(B)$ in $\sigma_{xx}(B)$.

The quantum--classical correspondence thus obtained provides useful knowledge to analyze the experimental data. 
Note that the classical formula has been used to analyze experimental data, even at strong fields, because it is useful, knowing that its validity will be lost for $\omega_c \tau \gtrsim 1$.
In contrast, the obtained quantum--classical correspondence guarantees the validity of the classical formulation, Eq. \eqref{Eq.1}, even for $\omega_c \tau\gtrsim 1$ except for the quantum oscillation.
Although this wide-range validity somehow contrasts with the conventional understanding, the present work clearly proves it both analytically and numerically.
In addition, the quantum-classical correspondence will also be useful to distill the quantum oscillation factor $Q(B)$ from the experimental data.

From the viewpoint of the analytic properties of thermal Green's function, an interesting perspective was present for the transport coefficient in the magnetic field.
As the field intensity increases, the primary constituent of Hall conductivity switches from the dissipative to dissipationless term. In contrast, only the dissipative term dominates among the diagonal components of the tensor.
The dissipationless term originates from the non-equilibrium transport which commonly gives rise to the thermodynamic phenomena like orbital diamagnetism \cite{Peierls1933,Fukuyama1970jpsj} and spin current in paramagnetic conductors \cite{Fuseya2015JPSJ}.
Magnetoresistance spontaneously includes both contributions.
Moreover, their weight is alternative depending on the intensity of the field and the relative direction of the current from the field.



Finally, the derived formula can be readily applied to multicarrier systems and quantum limit transport.
The peak or dip positions of quantum oscillation in certain metals and semimetals deviate from the intersection between the Landau level and the chemical potential.
The developed formulation demonstrated from a microscopic perspective that the origin was the dissipationless contribution to the off-diagonal conductivity, which superseded the dissipative terms in the diagonal conductivity in high-intensity fields.
Thus, we consider that the quantum extension derived herein provides new insights regarding quantum oscillation to the semiclassical quantitative magnetotransport theory.

Our study can also be extended to anisotropic electron systems such as ellipsoidal Fermi surfaces by altering the effective mass \cite{Ando1976anisotropic,Fuseya2015JPSJ}.
However, in multi-band systems, the interband contribution cannot always be renormalized to the mass.
In particular, we would need to consider the multi-band Hamiltonian and resulting energy dispersion and velocity operators, which is beyond the present theory.
In such cases, the quantum--classical correspondence is still unclear due to the non-trivial interband contribution \cite{Fuseya2015JPSJ,Fuseya2009}.

\begin{acknowledgements}
The authors would like to thank S. Tago for many helpful suggestions. We present special thanks to M. Tokunaga and A. Miyake for their valuable discussions. This work was supported by JSPS KAKENHI Grant Numbers 19H01850, 23H00268, and 23H04862.
\end{acknowledgements}


%

\end{document}